\documentclass[10pt,a4paper]{article}

\usepackage[margin=0.8in]{geometry}
\usepackage{amsmath,amssymb,amsfonts,accents,mathrsfs}
\usepackage{array}

\usepackage{graphicx}
\usepackage{textcomp}
\usepackage[svgnames]{xcolor} 

\definecolor{amethyst}{rgb}{0.6, 0.4, 0.8}

\usepackage[labelsep=quad,indention=10pt]{subfig}
\usepackage{algorithm}
\usepackage{algorithmic}

\usepackage{tikz}
\usetikzlibrary{positioning}
\usepackage{textcomp}
\usetikzlibrary{shapes,arrows,backgrounds}
\usetikzlibrary{datavisualization}
\usetikzlibrary{patterns}

\definecolor{x11_gray}{rgb}{0.85, 0.85, 0.85}

\def\BibTeX{{\rm B\kern-.05em{\sc i\kern-.025em b}\kern-.08em
    T\kern-.1667em\lower.7ex\hbox{E}\kern-.125emX}}

\usepackage{pgfplots}
\usepackage{tikz}
\usepgflibrary{plotmarks}
\usepgflibrary[plotmarks]
\usetikzlibrary{plotmarks}
\usetikzlibrary[plotmarks]
\usetikzlibrary{positioning}
\usepackage{textcomp}
\usetikzlibrary{shapes,arrows,backgrounds}
\usetikzlibrary{datavisualization}
\usetikzlibrary{patterns}
\pgfplotsset{compat=newest}
\pgfplotsset{plot coordinates/math parser=false}
\newlength\figureheight
\newlength\figurewidth

\pgfplotscreateplotcyclelist{list}{%
color=ForestGreen, solid, every mark/.append style={solid, fill=ForestGreen},line width=2.0pt, mark=*\\%
color=RoyalBlue,solid, every mark/.append style={solid, fill=RoyalBlue},line width=2.0pt, mark=square*\\%
color=Red,solid, every mark/.append style={solid, fill=Red},line width=2.0pt, mark=triangle*\\%
color=Goldenrod,solid, every mark/.append style={solid, fill=Goldenrod},line width=2.0pt,mark=diamond*\\%
color=DarkOrange,solid, every mark/.append style={solid, fill=DarkOrange},line width=2.0pt,mark=square*\\%
color=Purple,solid, every mark/.append style={solid, fill=Purple},line width=2.0pt,mark=otimes*\\%
color=LightSeaGreen,solid, every mark/.append style={solid, fill=LightSeaGreen},line width=2.0pt,mark=diamond*\\%
color=Black,solid, every mark/.append style={solid,fill=Black},line width=2.0pt\\%
color=Silver,solid,every mark/.append style={solid, fill=Silver}line width=2.0pt,,mark=diamond*\\%
}

\begin{document}

\begin{center}
\begin{LARGE}
Study of Activity-Aware Multiple Feedback Successive Interference
Cancellation for Massive Machine-Type Communications
\end{LARGE}

\vspace{10pt}
Roberto B. Di Renna and Rodrigo C. de Lamare \\
Center for Telecommunications Studies (CETUC) \\
Pontifical Catholic University of Rio de Janeiro, RJ, Brazil\\
E-mails: robertobrauer@cetuc.puc-rio.br, delamare@cetuc.puc-rio.br
\end{center}

\textit{\textbf{Abstract}} - In this work, we propose an activity-aware low-complexity multiple feedback successive interference cancellation (AA-MF-SIC) strategy for massive machine-type communications. The computational complexity of the proposed AA-MF-SIC is as low as the conventional SIC algorithm with very low additional complexity added. The simulation results show that the algorithm significantly outperforms the conventional SIC schemes and other proposals. \linebreak


\section{Introduction}

Massive machine-type communications (mMTC) have been considered as
one of the promising technologies in the future $5^{\textrm{th}}$
generation network. mMTC can be applied in many scenarios, including
IoT, smart cities, transportation communications and natural
disaster detection~\cite{HUAWEI2017}. Different from the
conventional human type communications, mMTC communications for IoT
have unique service features. Specifically, the unique features of
mMTC communications include the massive transmissions from a large
number of machine type communication devices (MTCDs), low data
rates, very short packets and high requirements of energy efficiency
and security~\cite{Tullberg2016},\cite{Chen2017},\cite{mmimo},
\cite{wence}.

One approach to reduce the overhead in sporadic mMTC is to avoid
control signalling regarding the activity of devices before
transmission. In this way, instead of the classical grant-based
random access scheme, the adoption of grant-free access schemes is
more promising. In grant-free access schemes, each packet is divided
in only two parts, preamble (metadata) and payload
(data)~\cite{Hasan2013}\cite{Liu2018}. The pilot sequence in each
packet metadata is used as the identification number (ID) of each
user in Code Division Multiple Access (CDMA) \cite{jpais}. This
sequence allows the base station (BS) to detect the active devices
and estimate their channels based on the received
metadata~\cite{Azari2017}. Thus, the BS can decode the data with the
estimated channels. As this scenario has a massive number of devices
with a low-activity probability, this problem can be interpreted as
a sparse signal processing problem.

As the standard problem changed, the detection algorithms should be
reformulated to this new scenario. In~\cite{Zhu2011} Zhu and
Giannakis proposed the Sparse Maximum a Posteriori Probability
(S-MAP) detection which, in a nutshell, performs a MAP detection of
the new sparse problem, considering the zero-augmented finite
alphabet. In the same paper, the authors proposed linear relaxed
S-MAP detectors, called Ridge (RD) and Lasso (LD) detectors.
Similarly to the classical Sphere Decoder, a sparsity-aware version
called K-Best has been proposed in~\cite{Knoop2014}.
In~\cite{Zhang2017} and~\cite{Zhang2018} the authors have proposed
solutions without knowing the activity factor $p_a$ (the probability
of active user). They belong to the class of Bayesian interference
algorithms and iterative reweighted approaches. Despite their good
performance, the computational complexity of these algorithms is
relatively large. In order to reduce this complexity,
in~\cite{Knoop2013} the sparsity-aware successive interference
cancellation (SA-SIC) has been presented. SA-SIC incorporates a
sparsity constraint into the detection process. Ahn et
al.~\cite{Ahn2018} reported a version of SA-SIC with a sorted
detection order, employing a sorted QR decomposition (SQRD) and an
alternative using the activity probability of devices (A-SQRD),
which outperforms SA-SIC.

Since in a massive machine-type communications scenarios we have
sporadically active devices and a requirement of low latency and
energy efficiency, the less retransmissions needed the better. In
this work, we propose an activity-aware multiple feedback
succcessive interference cancellation (AA-MF-SIC) technique to
reduce the error propagation of the SA-SIC algorithm. We draw
inspiration from a multi-feedback algorithm~\cite{Li2011} which
considers the feedback diversity by using a number of selected
constellation points as the feedback. Unlike prior work with
successive interference cancellation and list-based detectors
\cite{deLamare2003,itic,deLamare2008,cai2009,jiomimo,dfcc,deLamare2013,did,rrmser,bfidd,1bitidd},
AA-MF-SIC exploits the activity of devices. Using a smart
interference cancellation, a selection algorithm is introduced to
prevent the search space growing exponentially. As the alphabet has
changed, in the proposed AA-MF-SIC approach, the shadow constraints
are modified. Using the information of devices activity, different
constraints for each user are calculated. Simulation results show
that the proposed AA-MF-SIC scheme significantly outperforms the
literature schemes with a competitive complexity.

The organization of this paper is as follows:
Section~\ref{sec:System_Model} briefly describes the Low-Active CDMA
(LA-CDMA) system model and the augmented alphabet.
Section~\ref{sec:SA_SIC} introduces the Sparsity-MAP, Sparsity-Aware
SIC and the version with the A-SQRD algorithm while the
Section~\ref{sec:MF_SA_SIC} describes the conventional MF-SIC scheme
and the new AA-MF-SIC. Section~\ref{sec:Compl} compares the
complexity of each algorithm considered. Section~\ref{sec:Sim_res}
presents the set up for simulations and results while
Section~\ref{sec:Conc} draws the conclusions.

\textit{Notation:} Matrices and vectors are denoted by boldfaced
capital letters  and lower-case letters, respectively. The space of
complex (real) $N$-dimensional vectors is denoted by
$\mathbb{C}^N\left(\mathbb{R}^N\right)$. The $i$-th column of a
matrix $\mathbf{A} \in \mathbb{C}^{M\times N}$ is denoted by
$\mathbf{a}_i \in \mathbb{C}^M$. The superscripts
$\left(\cdot\right)^T$ and $\left(\cdot\right)^H$ stand for the
transpose and conjugate transpose, respectively, while
$\mathrm{tr}\left(\cdot\right)$ is the trace operator. For a given
vector $\mathbf{x} \in \mathbb{C}^N, ||\mathbf{x}||$ denotes its
Euclidean norm. $\mathbb{E\left[\cdot\right]}$ stands for expected
value and $\mathbf{I}$ is the identity matrix.

\section{System Model and Problem Statement}
\label{sec:System_Model}

The system model considered for the LA-CDMA uplink system consists
of $N$ MTC devices access a single base station, with a spreading
factor $M$ of the symbols for each device. At each time instant, the
system transmits $N$ symbols, taken from the constellation set
$\mathcal{A}$, organized into a column vector $\mathbf{x}$. The
symbol vector $\mathbf{x}$ is then transmitted over Rayleigh fading
channels, organized into a $M\times N$ channel matrix $\mathbf{H}$
which brings together the spreading sequences and channel impulse
responses to the base station. The received signal is collected into
a $M\times 1$ vector $\mathbf{y}$ given by
\begin{equation}\label{eq:main}
    \mathbf{y} = \mathbf{H}\mathbf{x} + \mathbf{n},
\end{equation}
where the $M \times 1$ vector $\mathbf{n}$ is a zero mean complex
circular symmetric Gaussian noise with covariance matrix
$\mathbb{E}\left[\mathbf{n}\mathbf{n}^H\right] =
\sigma_n^2\mathbf{I}$. The symbol vector $\mathbf{x}$ has zero mean
and covariance matrix $\mathbb{E}\left[\mathbf{x}\mathbf{x}^H\right]
= \sigma_x^2\mathbf{I}$, where $\sigma_x$ is the signal power. Each
symbol $x_n$ is drawn from the equi-probable finite alphabet
$\mathcal{A}$ when the $n$-th device is active, and zero otherwise.
So, considering a quadrature phase-shift keying (QPSK) modulation,
the augmented alphabet would be described as $\mathcal{A}_0 =
\mathcal{A} \cup \left\{0\right\}$, where $\mathcal{A} =
\left\{\left(1+j\right)/\sqrt{2},\left(-1+j\right)/\sqrt{2},\left(1-j\right)/\sqrt{2},\left(-1-j\right)/\sqrt{2}\right\}$.

As in this scenario devices have a low-activity probability, this
can be interpreted as a sparse signal processing problem. Most
algorithms proposed in the literature suggests the use of successive
interference cancellation, modified to detect the augmented
alphabet. In contrast, we propose a detection structure which, based
on the reliability of the previous estimates, can improve the
performance of the cancellation and reduce the need for block
retransmissions.



\section{Preliminary work}
\label{sec:prior}

In this section, we shall review existing interference cancellation
techniques for mMTC scenarios such as S-MAP, SA-SIC, SA-SIC with
A-SQRD and MF-SIC.

\subsection{Sparsity-MAP and SA-SIC Detections}

In order to detect $\mathbf{x}$ in (\ref{eq:main}), Zhu and
Giannakis\cite{Zhu2011} proposed a MAP detector that separates the
regularization constant which contains the activity probability
($p_n$) of each device. Since each entry $x_n$ is independent from
each other, the prior probability for $\mathbf{x}$ can be expressed
as

    \begin{eqnarray}\nonumber
        \mathrm{Pr}\left(\mathbf{x}\right) &=& \prod_{n=1}^{N} \mathrm{Pr}\left(x_n\right)\\ \label{eq:prod}
                                           &=& \prod_{n=1}^{N} \left(1-p_n\right)^{1-\left|x_n\right|_0}\left(p_n/\left|\mathcal{A}\right|\right)^{\left|x_n\right|_0} \\ \label{eq:prod2}
         \ln \mathrm{Pr}\left(\mathbf{x}\right) &=& - \ln \left(\frac{1-p_n}{p_n/\left|\mathcal{A}\right|}\right) \left|x_n\right|_0 + \ln \left(1 - p_n\right)
    \end{eqnarray}

where $\left|x_n\right|_0$ is the pseudo-norm that is zero if $x_n = 0$ (device is not active) or is one if $x_n \neq 0$. Substituting (\ref{eq:prod2}) in the output of the MAP detector, we obtain
    \begin{eqnarray}
        \mathbf{\hat{x}} &=& \underset{\mathbf{x}\in \mathcal{A}_0^N}{\textrm{arg max}}\hspace*{5pt} \mathrm{Pr}\left(\mathbf{x}|\mathbf{y}\right) \\ \label{eq:map}
                         &=& \underset{\mathbf{x}\in \mathcal{A}_0^N}{\textrm{arg min}}\hspace*{5pt} -\ln \mathrm{Pr}\left(\mathbf{y}|\mathbf{x}\right) - \ln \mathrm{Pr}\left(\mathbf{x}\right).
     \end{eqnarray}
 Introducing a regularization parameter given by
    \begin{eqnarray} \label{eq:lambda}
         \lambda_n       &=& \ln \left(\frac{1-p_n}{p_n/\left|\mathcal{A}\right|}\right)
    \end{eqnarray}
 \vspace{-2pt}
   we have the optimization problem which promotes the sparsity of $\mathbf{x}$ described by
    \begin{eqnarray}  \label{eq:smap}
        \mathbf{\hat{x}} &=& \underset{\mathbf{x}\in \mathcal{A}_0^N}{\textrm{arg min}}\hspace*{5pt} ||\mathbf{y} - \mathbf{H}\mathbf{x}||^2_2 + \lambda_n |x_n|_0.
    \end{eqnarray}

\vspace{-3pt}
Knoop in~\cite{Knoop2013} proposed a costless solution compared to the S-MAP, by introducing a successive interference cancellation into (\ref{eq:smap}). The original Sparsity-Aware Successive Interference Cancellation (SA-SIC) employs the QR decomposition but in order to have a fair comparison to our proposal, 
a version without the QR decomposition is considered. SA-SIC uses the regularization parameter to increase the reliability of the quantization process. This algorithm outperforms the conventional linear MMSE detector and gives a solution with much lower complexity than that of S-MAP. However, as SA-SIC does not order the  channel matrix columns before the cancellation, it is suitable to error propagation. Thus, appropriate detection order of channel matrix could increase the performance.

\vspace{-4pt}
\subsection{Sparsity-Aware SIC with Activity-Sorted QR Decomposition (SA-SIC with A-SQRD) Detection}
\label{sec:SA_SIC}

The proposed activity-aware sorted QR decomposition (A-SQRD) algorithm sorts the columns of the channel matrix $\mathbf{H}$ based on channel gains. The modification of the conventional SQRD algorithm is to include the regularization term $\lambda_n$ and the noise variance $\sigma^2_n$ in consideration in the detection ordering. Considering QPSK modulation, it is possible to replace the $l_o$-norm with the $l_2$-norm in (\ref{eq:smap}), since PSK constellations has a constant modulus alphabet ($||\mathbf{x}||_0 = ||\mathbf{x}||_2^2 = ||\mathbf{x}||_p^p, p \leq 1$)~\cite{Zhu2011}.
\vspace{-10pt}
    \begin{eqnarray} \nonumber
                    \mathbf{\hat{x}} &=& \underset{\mathbf{x}\in \mathcal{A}_0^N}{\textrm{arg min}}\hspace*{5pt} \left|\left|\mathbf{y} - \mathbf{H x}\right|\right|^2_2 + \sigma_n^2 \sum_{n=1}^N \lambda_n \left|x_n\right|^2 \\ \nonumber
                    &=& \underset{\mathbf{x}\in \mathcal{A}_0^N}{\textrm{arg min}}\hspace*{5pt} \left|\left|\mathbf{y} - \mathbf{H x}\right|\right|^2_2 + \left|\left| \sigma_n\, \textrm{diag}\left(\sqrt{\boldsymbol{\lambda}}\right) \mathbf{x}\right|\right|^2_2    \\ \nonumber
                    &=& \underset{\mathbf{x}\in \mathcal{A}_0^N}{\textrm{arg min}}\hspace*{5pt} \left|\left|\left[ \begin{array}{c} \mathbf{y} \\ \mathbf{0}_N  \end{array} \right] - \left[ \begin{array}{c} \mathbf{H} \\ \sigma_n\, \textrm{diag}\left(\sqrt{\boldsymbol{\lambda}}\right)    \end{array} \right] \mathbf{x}  \right|\right|^2_2 \\
                    &=& \underset{\mathbf{x}\in \mathcal{A}_0^N}{\textrm{arg min}}\hspace*{5pt} \left|\left|\mathbf{y}_0 - \mathbf{H' x}\right|\right|^2_2  \label{eq:zero-aug-obs}                     \end{eqnarray}

In order to find the best permutation of the columns of $\mathbf{H}'$, the A-SQRD algorithm employs also the modified Gram-Schmidt algorithm to reorder the columns of $\mathbf{H}'$ in (\ref{eq:zero-aug-obs}) before each orthogonalization step. Whereas our proposal does not uses the QR decomposition, a SA-SIC ordered by the channel norm with the Gram-Schmidt algorithm.

\section{Activity-Aware Multi-Feedback SIC detection (AA-MF-SIC)}
\label{sec:MF_SA_SIC}

This section is devoted to the description of the proposed
activity-aware multi-feedback successive interference cancellation
(AA-MF-SIC) detector for LA-CDMA systems. We present the overall
principles and structures of the proposed scheme in the first place,
and the we implement the proposed detector. Finally, we compare the
complexity of AA-MF-SIC with the existing techniques in the
literature.

The main idea of the AA-MF-SIC is to judge the reliability of each
estimated symbol, in a way to revisit other possible constellation
points if the previous estimate is not reliable. The reliability of
the previous detected symbol in the conventional MF-SIC is
determinate by the shadow area constraints (SAC). In this work, we
propose a modification to determine these constraints, saving
computational complexity by avoiding redundant processing.

In sequence, we describe the procedure for detecting $\hat{x}_n$ for
user $n$ so, other streams can be obtained accordingly. At each
cancellation step of the SIC cancellation process, described
in~(\ref{eq:sic}), the quantized estimated symbol $\hat{x}_n =
\mathcal{Q}\left[\mathbf{w}^H_n\mathbf{y}_n\right]$ is obtained
through the MMSE filter. In order to obtain accurate estimates, we
employ a an $l_1$-norm penalty function for sparse regularization in
the cost function, as described in the following optimization
problem:

    \begin{equation} \label{eq:cost_func}
    J\left(\mathbf{w}_n\right) = \mathrm{E}\left[||x_n - \hat{x}_n||^2_2\right] + 2\lambda_n||\mathbf{w}_n||_1
    \end{equation}

    \begin{eqnarray} \nonumber
        \underset{\mathbf{w}_n}{\textrm{min}}\hspace*{2.55pt} J\left(\mathbf{w}_n\right) &=& \underset{\mathbf{w}_n}{\textrm{arg min}}\hspace*{3pt} \mathrm{E}\left[||x_n - \mathbf{w}_n^\mathrm{H}\mathbf{y}_n||^2_2\right] + 2\lambda_n||\mathbf{w}_n||_1 \\ \nonumber
        &=& \underset{\mathbf{w}_n}{\textrm{arg min}}\hspace*{3pt} \mathrm{E}\left[\mathrm{tr}\left\{\left(x_n - \mathbf{w}_n^\mathrm{H}\mathbf{y}_n\right)^\mathrm{H}\right. \right.\\
        && \hspace{30pt} \left.\left.\left(x_n - \mathbf{w}_n^\mathrm{H}\mathbf{y}_n\right)\right\}\right]+\, 2\lambda||\mathbf{w}_n||_1
    \end{eqnarray}

The use of the $l_1$-norm penalty function imposes some difficulty
as the cost function $J\left(\mathbf{w}_n\right)$ is
non-differentiable. Zhu~\cite{Zhu2011} considers the use of
quadratic programming solvers but in a way to reduce the
computational complexity, we make an approximation to the
regularization term \cite{Angelosante2010,l1stap,saalt}, which is
given by,
    \begin{equation}\label{eq:approx_w}
        ||\mathbf{w}_n||_1 \approx \mathbf{w}_n^\mathrm{H} \Lambda \mathbf{w}_n
    \end{equation}
where
    \begin{equation}
        \Lambda = \mathrm{diag}\left\{\frac{1}{|w_{n,1}|+\epsilon},\frac{1}{|w_{n,2}|+\epsilon},\cdots,\frac{1}{|w_\mathrm{n,M}|+\epsilon}\right\}
    \end{equation}

and $\epsilon$ is a small positive constant. Fixing the term
$\Lambda$, we take the derivative of (\ref{eq:approx_w}) with
respect to $\mathbf{w}^\ast$, with the knowledge of
\begin{equation}
        \frac{\partial ||\mathbf{w}_n||_1}{\partial\mathbf{w}_n^\ast} \approx \Lambda \mathbf{w}_n
\end{equation}
\begin{equation}
        \nabla J\left(\mathbf{w}_n\right)_{\mathbf{w}_n^\ast}  = \mathrm{E}\left[-\mathbf{y}_n\left(x_n - \mathbf{w}_n^\mathrm{H}\mathbf{y}_n\right)^\mathrm{H}\right] + 2\lambda_n \Lambda \mathbf{w}_n
\end{equation}

By equating the above equation to a zero vector, we obtain the
filter weight vector:

        \begin{eqnarray} \nonumber
            \mathbf{0}_{\mathrm{M}\times 1} &=& -\mathrm{E}\left[\mathbf{y}_n x_n^\mathrm{H}\right] + \mathbf{w}_n\mathrm{E}\left[\left(\mathbf{y}_n\mathbf{y}_n^\mathrm{H}\right)\right] + 2\lambda_n\Lambda \mathbf{w}_n  \\ \nonumber
            &=& -\sigma_x^2\mathbf{H}'\boldsymbol{\delta}_n + \mathbf{w}_n\left(\sigma^2_x\mathbf{H}'\mathbf{H}^\mathrm{'H} +{\sigma_n^2}\mathbf{I}\right)+ 2\lambda_n\Lambda \mathbf{w}_n  \\
            \mathbf{w}_n &=&  \left(\overline{\mathbf{H}}_n\overline{\mathbf{H}}_n^\mathrm{H} +\frac{\sigma_n^2}{\sigma_x^2}\mathbf{I} + \frac{2\lambda_n}{\sigma_x^2}\Lambda\right)^{-1}   \overline{\mathbf{H}}_n\boldsymbol{\delta}_n
        \end{eqnarray}

        where $\boldsymbol{\delta}_n$ is a $\mathrm{N} \times 1$ zero column vector with 1 at the $n$-th position. $\overline{\mathbf{H}}_n$ denotes the matrix obtained by taking the columns $n, n+1, \dots , N$ of $\mathbf{H}'$ and $\mathcal{Q}\left[\cdot\right]$ is the quantization function appropriate for the modulation scheme being used in the system. The quantization operates by choosing the constellation point with the smallest Euclidean distance to the estimated symbol. The SIC is carried out as follows:

    \begin{equation}
        \begin{array}{llll}
        \mathbf{y}_n &=& \mathbf{y}, & n=1,\\ \label{eq:sic}
        \mathbf{y}_n &=& \mathbf{y} - \sum^{n-1}_{j=1} \mathbf{h}'_j \hat{x}_j, & n\geq 2
        \end{array}
    \end{equation}
where $n$ refers to the cancellation stage. This modified filter is
an iterative version of the MMSE filter, as $\Lambda$ depends on
previous estimates. At each new stream the filter is updated, using
the corresponding activity probability regularization ($\lambda_n$)
after the SIC operation.

\subsection{Shadow Area Constraints}

\begin{figure}
  \centering
        \subfloat[Conventional MF-SIC.]{
            \begin{tikzpicture}[
            scale=0.45,
            axis/.style={very thick, ->, >=stealth'},
            important line/.style={thick},
            dashed line/.style={dashed, thin},
            pile/.style={thick, ->, >=stealth', shorten <=2pt, shorten
            >=2pt},
            every node/.style={color=black}
            ]
            \draw[axis] (-6.5,-2)  -- (2.5,-2) node(xline)[above]
                {$I$};
            \draw[axis] (-2,-6.5) -- (-2,2.5) node(yline)[right] {$Q$};

            \foreach \x in {-4,0}
            \foreach \y in {-4,0}
            {
            \draw[dashed] (\x,\y) circle (1.2cm);
            \fill (\x,\y) circle (0.1cm);
            }

            \begin{scope}
                \fill[fill=x11_gray] (-6,-1.95) rectangle (-2.05,2) (-4,0) circle (1.2cm);
                \fill[fill=x11_gray] (-6,-6) rectangle (-2.05,-2.05) (-4,-4) circle (1.2cm);
                \fill[fill=x11_gray] (-1.95,-1.95) rectangle (2,2) (0,0) circle (1.2cm);
                \fill[fill=x11_gray] (-1.95,-6) rectangle (2,-2.05) (0,-4) circle (1.2cm);
            \end{scope}


            \node (dth_c) at (-3.7, -0.2) {};
            \node (dth_r) at (-5.3, 0.7) {};
            \node (uk_c) at (-3.7, 0.3) {};
            \node (uk_r) at (-6, -2) {};
            \node (zn) at (-5.0,-1.7) {{\footnotesize{$z_{n}$}}};
            \filldraw[fill=gray!80!white] (-5.7,-1.7) circle (0.1cm);
            \node (dk) at (-5.45,-0.8) {{\footnotesize{$d^{k}$}}};
            \node (dth) at (-3.9,0.6) {{\footnotesize{$d^{th}$}}};

            \node (a4) at (-3.6,-0.4) {{\footnotesize{$a_{4}$}}};
            \node (a1) at (0.4,-0.4) {{\footnotesize{$a_{1}$}}};
            \node (a3) at (-3.6,-4.4) {{\footnotesize{$a_{3}$}}};
            \node (a2) at (0.4,-4.4) {{\footnotesize{$a_{2}$}}};

            \draw[->,>=stealth] (dth_c) edge (dth_r);
            \draw[->,>=stealth] (uk_c) edge (uk_r);
            \end{tikzpicture}
    \label{fig:Conv_MF}
    }
    \subfloat[MF-SIC with Activity information.]{
    \centering
            \begin{tikzpicture}[
            scale=0.45,
            axis/.style={very thick, ->, >=stealth'},
            important line/.style={thick},
            dashed line/.style={dashed, thin},
            pile/.style={thick, ->, >=stealth', shorten <=2pt, shorten
            >=2pt},
            every node/.style={color=black}
            ]
            \draw[axis] (-6.5,-2)  -- (2.5,-2) node(xline)[above]
                {$I$};
            \draw[axis] (-2,-6.5) -- (-2,2.5) node(yline)[right] {$Q$};

            \foreach \x in {-4,0}
            \foreach \y in {-4,0}
            {
            \draw[dashed] (\x,\y) circle (1.2cm);
            \fill (\x,\y) circle (0.1cm);
            }

            \begin{scope}
                \fill[fill=x11_gray] (-6,-1.95) rectangle (-2.05,2) (-4,0) circle (1.2cm);
                \fill[fill=x11_gray] (-6,-6) rectangle (-2.05,-2.05) (-4,-4) circle (1.2cm);
                \fill[fill=x11_gray] (-1.95,-1.95) rectangle (2,2) (0,0) circle (1.2cm);
                \fill[fill=x11_gray] (-1.95,-6) rectangle (2,-2.05) (0,-4) circle (1.2cm);
                \draw[fill=white,dashed] (-2,-2) circle (1cm);
            \end{scope}

            \fill (-2,-2) circle (0.1cm);

            \node (dth_c) at (-3.7, -0.2) {};
            \node (dth_r) at (-5.3, 0.7) {};
            \node (uk_c) at (-3.7, 0.3) {};
            \node (uk_r) at (-6, -2) {};
            \node (dth_c0) at (-1.8, -2.15) {};
            \node (dth_r0) at (-3.1, -1.2) {};
            \node (zn) at (-5.0,-1.7) {{\footnotesize{$z_{n}$}}};
            \filldraw[fill=gray!80!white] (-5.7,-1.7) circle (0.1cm);
            \node (dk) at (-5.45,-0.8) {{\footnotesize{$d^{k}$}}};
            \node (dth) at (-3.9,0.6) {{\footnotesize{$d^{th}$}}};
            \node (dth0) at (-1.8,-1.4) {{\footnotesize{$d^{{th}_{0}}$}}};

            \node (a0) at (-1.6,-2.4) {{\footnotesize{$a_{0}$}}};
            \node (a4) at (-3.6,-0.4) {{\footnotesize{$a_{4}$}}};
            \node (a1) at (0.4,-0.4) {{\footnotesize{$a_{1}$}}};
            \node (a3) at (-3.6,-4.4) {{\footnotesize{$a_{3}$}}};
            \node (a2) at (0.4,-4.4) {{\footnotesize{$a_{2}$}}};

            \draw[->,>=stealth] (dth_c) edge (dth_r);
            \draw[->,>=stealth] (dth_c0) edge (dth_r0);
            \draw[->,>=stealth] (uk_c) edge (uk_r);
            \end{tikzpicture}
        \label{fig:MF_Act}
    }
    \caption{QPSK Constellations difference.}
    \label{fig:constellations}
\end{figure}
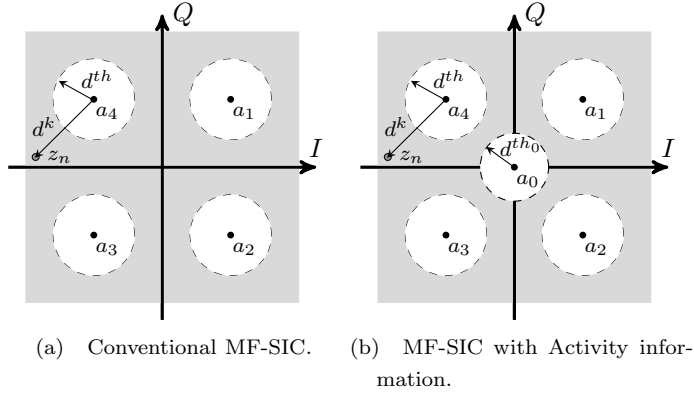

The SAC tests, for each user, the reliability of the result of the
soft estimate. The radius of reliability ($d^{th}$) is used to
determine the probability of the soft decision $z_n =
\mathbf{w}^H_n\mathbf{y}_n$ to drop into the shadow area on the
constellation  map. In the conventional MF-SIC, a radius of
reliability ($d^{th}$) is determined by successive tests and if the
nearest distance between the estimation and the constellation points
($a_f$) $d^k$, is higher than $d^{th}$, the estimation is considered
unreliable. The shadow area and those quantities are represented in
Fig.~\ref{fig:Conv_MF} and the radius of reliability is expressed by

\vspace{-14pt}
    \begin{eqnarray} \label{eq:dk}
        d^k &=& \underset{a_f\in \mathcal{A}_0}{\textrm{arg min}}\hspace*{5pt} |z_n - a_f|.
    \end{eqnarray}


\begin{figure}[H]
        \begin{center}
            \includegraphics[scale=.6]{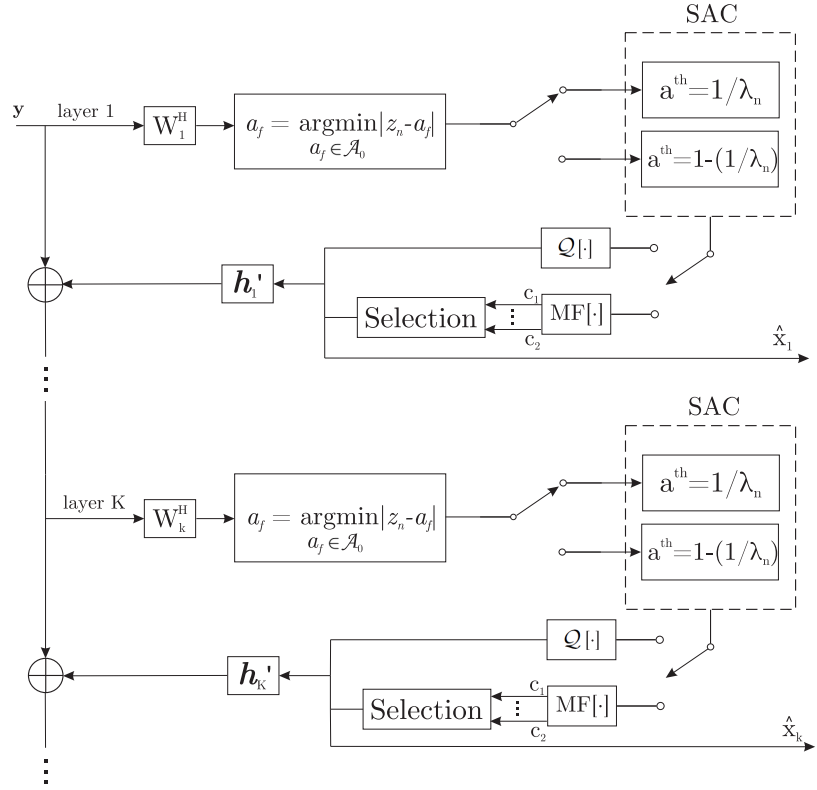}
            \caption{AA-MF-SIC scheme. The modified SAC determines the reliability of the filter output, using the regularization parameter.}
            \label{fig:sac}
        \end{center}
    \end{figure}

As in the mMTC scenarios is considered an augmented alphabet, SAC
also changes. Fig.~\ref{fig:MF_Act} shows the constellation diagram
with different constraints of the augmented QPSK alphabet. Since
those elements do not have the same \textit{a priori} probabilities,
those constraints can be determined by the information of devices
activity, $\lambda_n$. In the proposed SAC, after we compute all the
$d^k$s, if the closest point of the augmented alphabet to the soft
estimate is zero, $d^k$ is compared to $1/\lambda_n$, as shown in
Fig.~\ref{fig:sac}. Otherwise, the comparison is made with the
complement value of the regularization parameter, $d^{th} =
\left(1-1/\lambda_n\right)$. Therefore, we can see from the above
equations that, a larger distance $d^{th}$ corresponds to a higher
chance of making $z_n$ unreliable. If in the conventional MF-SIC the
radius of reliability is determinate by successive tests, here we
use the probability to being active of each device to define it.
%

    \begin{figure}[H]
        \begin{center}
            \includegraphics[scale=.95]{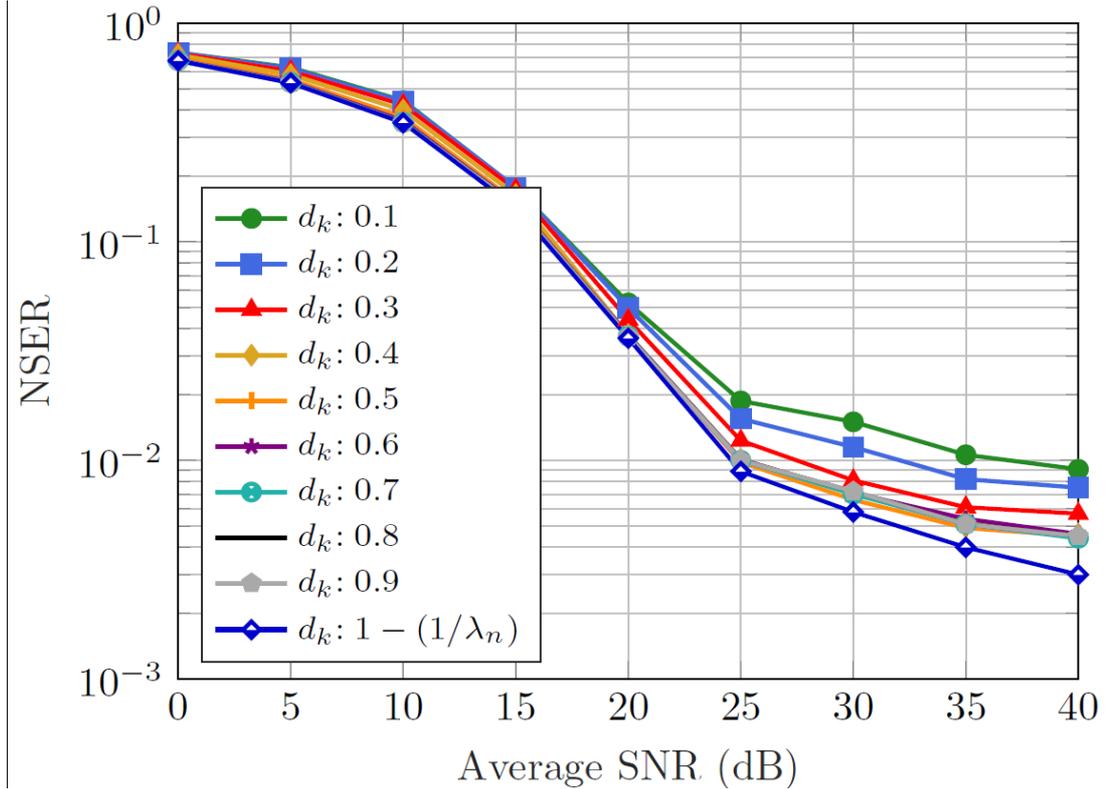}
            \caption{NSER vs Activity probability in different SNR values and perfect CSI.}
            \label{fig:results4}
        \end{center}
    \end{figure}

\subsection{Optimal feedback selection}

After the SAC assessment, if the soft estimate $z_n$ is considered
reliable, the algorithm ignores the MF approach and proceeds with
the conventional SIC scheme, as in~(\ref{eq:sic}). On the other
hand, if $z_n$ is considered unreliable, a candidate vector called
the MF set $\mathbf{b}$ is generated. The candidate vector is
composed by $F$ constellation points with minimum Euclidean distance
to $z_n$. Naturally, in this scenario, as the probability of not
being active is higher then being active, the zero is always
included in the MF set. The size of the MF set can be flexible or
predefined. The higher SNR corresponds to a smaller MF set size
which introduces a trade off between the complexity and the
performance. After the MF procedure, $\hat{x}_n$ is selected between
one of the candidates. With this approach, as the chosen
augmented-alphabet was QPSK, all possible symbols were considered.
All the benefits provided by the AA-MF-SIC algorithm are based on
the assumption that the optimal feedback candidate is efficiently
selected.

The algorithm starts with the definition of a set of $N \times 1$
vectors $\mathbf{b}_f$ which constitutes part of the $N \times F$
matrix $\mathbf{B}$. Each $\mathbf{b}_f$ is obtained by using the
candidate as the cancellation symbol in the $n-th$ device and
successively processing the remaining $n+1$ to $N$ devices with the
conventional SIC scheme resulting in the following $\mathbf{B}$
matrix:

    \begin{equation}\label{eq:c_mf}
        \mathbf{B} = \left[
        \begin{array}{cccc}
            \hat{x}_{1,1}     & \hat{x}_{1,2}     & \cdots & \hat{x}_{1,F} \\
            \hat{x}_{2,1}     & \hat{x}_{2,2}     & \cdots & \hat{x}_{2,F} \\
            \vdots            & \vdots            & \ddots & \vdots \\
            \hat{x}_{{n-1},1} & \hat{x}_{{n-1},2} & \cdots & \hat{x}_{{n-1},F} \\
            c_{1}             & c_{2}             & \cdots & c_{F} \\
            b_{i,1}           & b_{i,2}           & \cdots & b_{i,F} \\
            \vdots            & \vdots            & \ddots & \vdots \\
            b_{N,1}           & b_{N,2}           & \cdots & b_{N,F} \\
        \end{array} \right]
    \end{equation}

where the $\hat{x}_1, \dots, \hat{x}_{n-1}$ are previously detected
symbols and $c_1, \dots, c_{F}$ are the candidate symbols. The rest
of the elements are obtained by

    \begin{equation}\label{eq:rest_ele}
        b_{i,f} = \mathcal{Q}\left[\mathbf{w}^H_{i+1}\mathbf{y}^{\textrm{MF}}_{f}\right]
    \end{equation}

where ${\bf Y}^{\rm MF} = \left[{\bf y}^{\rm MF}_{1} \ldots \, {\bf
y}^{\rm MF}_{\textrm{F}}\right]$. For each candidate, each column of
$\mathbf{Y}^{MF}$ is updated as

    \begin{equation}
        \mathbf{y}^{\textrm{MF}}_f[i] = \mathbf{y}^{\textrm{MF}}_f[i-1] - \sum_{i=j}^{N-1} b_{i,f}\mathbf{h}'_i
    \end{equation}
The same MMSE filter is used for all the feedback candidates and
devices, which allows the proposed AA-MF-SIC algorithm
\cite{aamfsic} to have the computational simplicity of the
conventional SIC detection. Other approaches to compute the MMSE
filter such as \cite{jiomimo,jidf} can be considered. The proposed
AA-MF-SIC algorithm selects the candidate according to

    \begin{equation}
        \mathbf{b}^\textrm{opt} =\underset{f\leftarrow1,\cdots,F}{\textrm{arg min}}\hspace*{5pt} ||\mathbf{y} - \mathbf{H}' \mathbf{b}_f||^2
    \end{equation}

The optimum candidate, $b^\textrm{opt}_{j}$, is chosen to be the
optimal feedback symbol for the next user as well as a more reliable
decision for the current user ($\hat{x}_j = b^\textrm{opt}_{j}$).
The algorithm of the proposed AA-MF-SIC is summarized in
Algorithm~\ref{alg:AAMFSIC}.
%

\begin{algorithm}[t]
\footnotesize
\caption{AA-MF-SIC}
    \begin{algorithmic}[1] \label{alg:AAMFSIC}
        \REQUIRE $\mathbf{y}$, $\mathbf{H}$, $\mathcal{A}$, $\mathcal{A}_0$, $\sigma^2_w$, $\left\{p_n\right\}^{N}_{n=1}$, $\mathbf{H}'$, $\mathbf{q}$, $\boldsymbol{\lambda}$,$\epsilon$
        \ENSURE $\mathbf{\hat{x}}$ \\
        \textbf{\% Initialization and filtering}
        \STATE  $F \leftarrow \textrm{length}\left(\mathcal{A}_0\right)$; $\mathbf{R} \leftarrow \mathbf{H}'\mathbf{H}'^H + \sigma_w^2 \mathbf{I}$; $\mathbf{W} \leftarrow \mathbf{H}'\mathbf{R}^{-1}$\\
        \textbf{\% Detection}
        \FOR {$k \leftarrow 0,\cdots , M-1$}
        \STATE $\mathbf{H}^{\textrm{MF}} \leftarrow \mathbf{H}'$; $\mathbf{y}^{\textrm{MF}} \leftarrow \mathbf{y}$\\
        \textbf{\% SIC operation}
        \STATE $\mathbf{B} \leftarrow \textrm{zeros}\left[N,F\right]$
        \FOR{$j \leftarrow 1, \cdots, N$}
        \STATE $z_{j,{k+1}} \leftarrow \mathbf{w}_j^{\textrm{H}}\mathbf{y}^{\textrm{MF}}$; $\hat{x}_{j} \leftarrow \mathcal{Q}\left[z_{j,{k+1}}\right]$
        \STATE $b_{j} \leftarrow  \hat{x}_{j}$ \% all columns of matrix $\mathbf{B}$ in line $j$ receives $\hat{x}_{j}$
        \STATE $a_{\textrm{opt}} \leftarrow \underset{s\leftarrow1,\cdots,F}{\textrm{arg min}}\hspace*{5pt} |z_{j,{k+1}} - a_s|$\\
        \textbf{\% Multi-Feedback}
        \IF { $a_{\textrm{opt}}$ is related to zero}
        \STATE $d^{th}_j \leftarrow 1/\lambda_j $
        \ELSE
        \STATE $d^{th}_j \leftarrow 1-\left(1/\lambda_j\right) $
        \ENDIF
        \IF {$\left|\Re\left[z_{j,k+1}\right]\right| \textrm{and} \left|\Im\left[z_{j,k+1}\right]\right|  >  d^{th}_{j}$}
        \STATE Fill the columns of $\mathbf{Y}^{\textrm{MF}}$ matrix with F repetitions of $\mathbf{y}^{\textrm{MF}}$
        \FOR {$i \leftarrow j,\cdots , N-1$}
        \FOR {$f \leftarrow 1,\cdots , F$}
        \STATE $\mathbf{y}^{\textrm{MF}}_{f} \leftarrow \mathbf{y}^{\textrm{MF}}_{f} - b_{i,f} \mathbf{h}_i'$
        \STATE $b_{i+1,f} \leftarrow \mathcal{Q}\left[\mathbf{w}_{i+1}^H \mathbf{y}^{\textrm{MF}}_{f}\right]$
        \ENDFOR
        \ENDFOR
        \ENDIF
        \STATE $\mathbf{b}^\textrm{opt} \leftarrow \underset{f\leftarrow1,\cdots,F}{\textrm{arg min}}\hspace*{5pt} ||\mathbf{y} - \mathbf{H}' \mathbf{b}_f||^2$
        \STATE $\hat{x}_j \leftarrow b^\textrm{opt}_{j}$\\
        \textbf{\% Interference cancellation}
        \STATE $\mathbf{y}^{\textrm{MF}} \leftarrow \mathbf{y}^{\textrm{MF}} - \mathbf{h}^{\textrm{MF}}_j\hat{x}_j$
        \STATE $\mathbf{h}^{\textrm{MF}}_j \leftarrow \textrm{zeros}\left[M,1\right]$
        \STATE $\Lambda = \mathrm{diag}\left\{\frac{1}{|w_{1,j}|+\epsilon},\frac{1}{|w_{2,j}|+\epsilon},\cdots,\frac{1}{|w_{\mathrm{M},j}|+\epsilon}\right\}$
        \STATE $\mathbf{w}_j \rightarrow \left(\mathbf{H}^\mathrm{MF}\mathbf{H}^\mathrm{MF\, H} +\frac{\sigma_n^2}{\sigma_x^2}\mathbf{I} + \frac{2\lambda_n}{\sigma_x^2}\Lambda\right)^{-1}\mathbf{H}^\mathrm{MF}\boldsymbol{\delta}_j$
        \ENDFOR
        \ENDFOR
        \STATE $\hat{\mathbf{x}} \leftarrow \hat{\mathbf{x}}^T $; Reordering $\mathbf{\hat{x}}$ with $\mathbf{q}$
    \end{algorithmic}
\end{algorithm}

\section{Complexity Analysis}
\label{sec:Compl}

The detailed computational complexity is shown in terms of the
average number of required complex multiplications per symbol
detection. Considering $N$ as the number of devices and $M$ the
spreading gain and the linear MMSE as a lower bound, is possible to
compare complexities of all the simulated algorithms.
Table~\ref{tab:complexity} show that the Iterative Reweighted (IR)
algorithm has a lower complexity than the existing non-linear
schemes, but has a $L$ factor which means the number of iterations
required to converge to a refined estimation. As the K-Best is a
variant of the Sphere Decoder and depends on $K$, which is the
number of paths required to minimize the sum of per-symbol cost
functions, it is clearly the algorithm which requires more
computational effort. There is a slight difference between SA-SIC,
ordered SA-SIC and SA-SIC with A-SQRD, which is the type of ordering
of the channel matrix. A-SQRD performs the QR decomposition of the
augmented channel matrix and the Gram-Schmidt algorithm while
ordered SA-SIC just do the channel norm ordering. With an
intermediate complexity, AA-MF-SIC depends on the SNR to calculate
the required multiplications. As the usage of the SAC verifies the
reliability of the soft estimations, low SNRs demands more
multiplications than at high SNRs.

\begin{table}[t]
    \begin{center}
    \caption{Complexity algorithms comparison.}
        \begin{tabular}{ll} \hline
            Algorithm & Required complex multiplications \\ \hline
            MMSE & $3N^2 + N +1$\\
            IR & $L\left(3N^2 + N +1\right)$\\
            SA-SIC & $\left(1/6\right)\left(3N^3 + 11N^2 + 21N -2\right)$\\
            K-Best & $K \left|\mathcal{A}_{0}\right| \left(\frac{N^3}{3}+2N^2 + \frac{5}{3}N+\log^2\left(K\left|\mathcal{A}_0\right|\right)\right)$\\
            Ordered SA-SIC &$\left(1/6\right)\left(3N^3 + 11N^2 + 21N -2\right)$\\
            SA-SIC with A-SQRD & $2N^3+\left(2M+2\right)N^2 + \left(M-1\right)N$\\
            AA-MF-SIC high SNR & $\leq \left(1/6\right)\left(3N^3 + 11N^2 +21N -2\right)$ \\
            AA-MF-SIC low SNR & $\geq \left(1/6\right)\left(3N^3 + 11N^2 +21N -2\right)+10N^2$ \\   \hline
        \end{tabular}
        \label{tab:complexity}
    \end{center}
\end{table}

\section{Simulation Results}
\label{sec:Sim_res}

In this section, two different scenarios of a LA-CDMA uplink
communication system are considered, one with perfect channel state
information (CSI) at the receiver and a second one considering an
imperfect CSI scenario. We evaluate the symbol error rate of active
devices (Net Symbol Error Rate, NSER) performance of the proposed
AA-MF-SIC algorithm in an uncoded block fading channel system. NSER
performance of the proposed AA-MF-SIC is compared to MMSE, SA-SIC,
Iterative Reweighed (IR), K-Best, ordered SA-SIC and SA-SIC with
A-SQRD detectors. We note that coded systems with Low-Density
Parity-Check Codes (LDPC) \cite{dopeg,memd,lrbsc} can also be
considered.

Considering the AA-MF-SIC and all their counterparts in the
independent and identically-distributed (i.i.d.) random flat fading
model, where the coefficients are taken from complex Gaussian random
variables with zero mean and unit variance. Thus, the average SNR is
set to $1/\sigma^2_w$. At the transmitter end, all the antennas
(when the device was active) radiate QPSK symbols with the same
power. In the following experiments, we average the curves over
10000 runs. The receive processing is performed under the MMSE
criterion.

In order to provide a fair comparison with the results of the
algorithms in the literature, the simulation setup was the same as
in~\cite{Ahn2018}. The under-determined mMTC system simulated
considered has 128 (N) devices and a length of 64 (M) for spreading.
The activity probabilities are $\left\{p_n\right\}^N_{n=1}$ drawn
uniformly at random in $\left[0.1, 0.3\right]$.

\begin{figure}[H]
    \begin{center}
        \includegraphics[scale=.7]{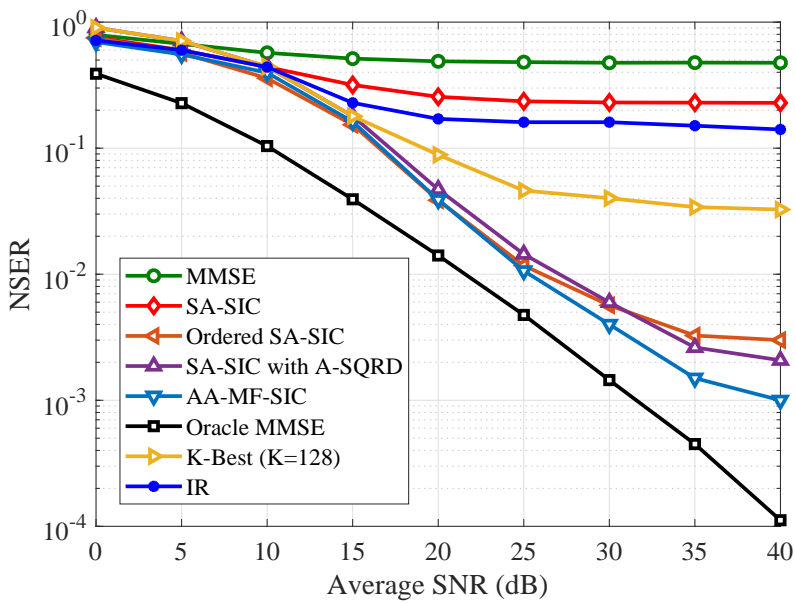}
        \caption{NSER vs. Average SNR with augmented QPSK modulation uncoded system and perfect CSI.}
        \label{fig:results}
    \end{center}
\end{figure}

Fig.~\ref{fig:results} shows the NSER performance for each algorithm as a function of the average SNR. The curves shows that the proposed AA-MF-SIC outperforms the literature algorithms. The modified versions of SA-SIC (without QR decomposition) and the proposed SA-SIC with A-SQRD as in~\cite{Ahn2018} are considered. The ``Ordered SA-SIC'' uses the channel norm sort. The lower bound Oracle MMSE is the traditional MMSE filter but with the aid of the information of which device is active or not.

\begin{figure}[H]
        \begin{center}
            \includegraphics[scale=.7]{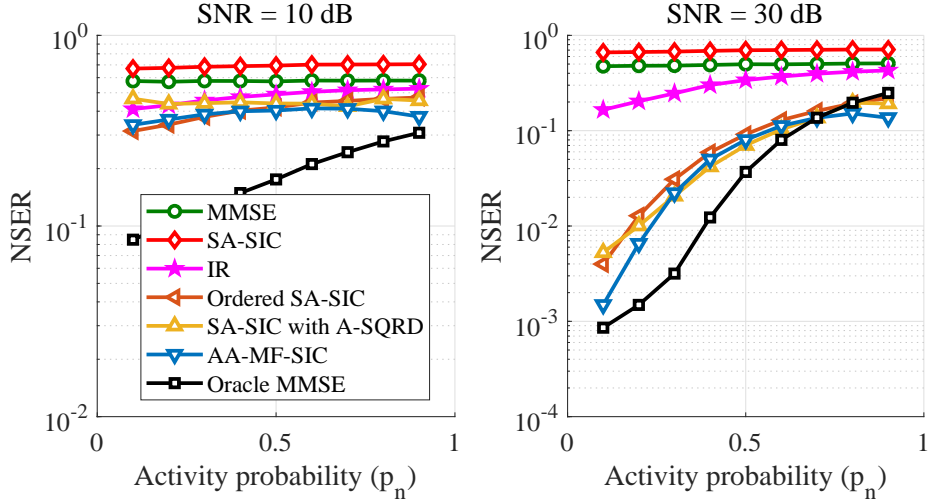}
            \caption{NSER vs Activity probability in different SNR values and perfect CSI.}
            \label{fig:results2}
        \end{center}
    \end{figure}

For the sake of verifying the behaviour of the algorithms if the probability of activity increases, other scenarios were considered. Fig.~\ref{fig:results2} presents the NSER performance of the algorithms with different fixed average SNR values. We notice that even when the activity probability is closer to 1, AA-MF-SIC has a satisfactory performance, becoming better than Oracle MMSE at higher $p_n$ values.

\begin{figure}[H]
    \begin{center}
        \includegraphics[scale=.7]{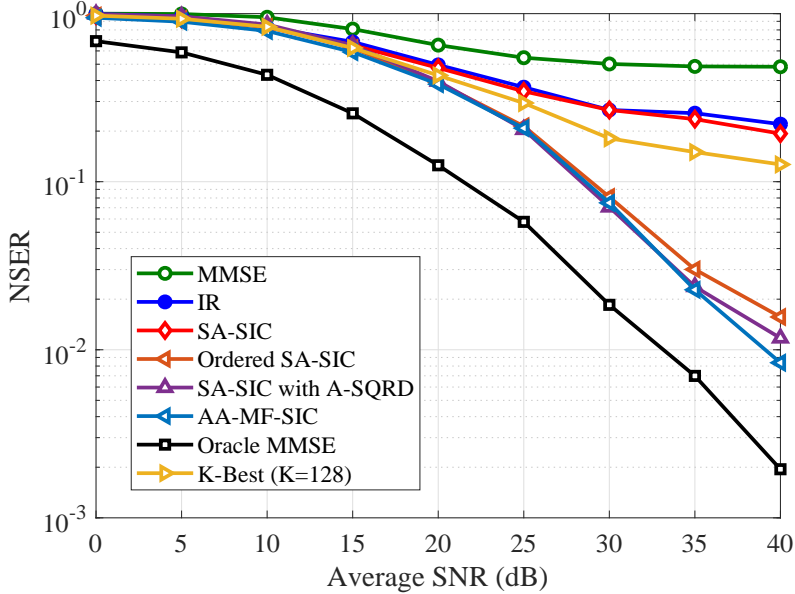}
        \caption{NSER vs. Average SNR with augmented QPSK modulation uncoded system and imperfect CSI.}
        \label{fig:results3}
    \end{center}
\end{figure}

Due to the sparsity feature of the transmitted signal $\mathbf{x}$, conventional channel estimators as LS and RLS do not work properly. As this is an open problem, considering estimation errors, the channel can be written as
\vspace{-3pt}
\begin{equation}\label{eq:channel_est}
    \hat{\bf H} = {\bf H} + {\bf E}
\end{equation}

where $\mathbf{H}$ represents the channel estimate and ${\bf E}$ is a random matrix corresponding to the error for each link. The channel for user $k$ can be written as $\hat{\bf h}_k = {\bf h}_k + {\bf e}_k$. Each coefficient of the error matrix follows a Gaussian distribution, i.e.,$\sim \mathcal{C}\mathcal{N}\left(0, \hat{\sigma}^2\right)$. Fig.~\ref{fig:results3} compares the performance of the considered algorithms.

\section{Conclusions}
\label{sec:Conc}

In this paper, we have considered the design of a low-complexity
detection algorithm for mMTC scenarios. In this context, compared
with previous works, we have presented an algorithm to mitigate
error propagation by using a multi-feedback aided successive
interference cancellation detection with modified shadow area
constraints. This approach effectively reduces error propagation in
decision driven interference cancellation techniques while
maintaining the low complexity of the already proposed algorithms.
The proposed scheme has been demonstrated to improve the performance
of existing algorithms, even in a scenario with higher activity
probability.

\end{document}